\documentclass[10pt]{article}
\usepackage{graphicx} 

\usepackage[whole]{bxcjkjatype} 

\usepackage[utf8]{inputenc} 
\usepackage[T1]{fontenc}    
\usepackage{hyperref}       
\usepackage{url}            
\usepackage{booktabs}       
\usepackage{amsfonts}       
\usepackage{nicefrac}       
\usepackage{microtype}      
\usepackage{xcolor}         
\usepackage{graphicx}       
\usepackage{amsmath}        
\usepackage{amsthm}
\usepackage{amssymb}        
\usepackage{wrapfig}
\usepackage{subcaption}

\usepackage{algorithm}
\usepackage{algorithmic}
\usepackage{natbib}

\title{Beyond Individuals: Collective Predictive Coding for Memory, Attention, and the Emergence of Language}
\author{Tadahiro Taniguchi\\ Graduate School of Informatics \\ Kyoto University\\ taniguchi@i.kyoto-u.ac.jp}
\date{}

\begin{document}
\maketitle

\begin{abstract}
This commentary extends the discussion by Parr et al. on memory and attention beyond individual cognitive systems. From the perspective of the Collective Predictive Coding (CPC) hypothesis---a framework for understanding these faculties and the emergence of language at the group level---we introduce a hypothetical idea: that language, with its embedded distributional semantics, serves as a collectively formed external representation. CPC generalises the concepts of individual memory and attention to the collective level. This offers a new perspective on how shared linguistic structures, which may embrace collective world models learned through next-word prediction, emerge from and shape group-level cognition.
\end{abstract}

\section{Introduction}
Language is formed collectively. Each individual cognitive system participating in its emergence possesses memory and attention. These faculties are undoubtedly crucial not only for modeling and understanding language but also for its formation, or emergence.

Language possesses a semantic-syntactic structure, and modeling its sequential nature certainly yields significant information. As \citet{Parr2025beyond} intuitively describe in their contribution to this special issue, mechanisms for storing such information (memory) and for retrieving information therefrom (attention) play vital roles in extracting information from long contexts and sequences with internal structures beyond Markov. Attention mechanisms in neural networks gained prominence through Transformers \citep{Vaswani2017attention} and have been applied to many artificial cognitive systems. Meanwhile, for long-sequence memory, State Space Models like Mamba \citep{Gu2023mamba} have garnered considerable attention.

\section{Main Discussion}

The importance of embedding spaces, as highlighted by \citet{Parr2025beyond}, is also well-taken. This concept is rooted in linguistic structures that give rise to distributional semantics \citep{Harris1954distributional}.
However, a crucial question arises: who embedded the ``retrievable'' information and structures into the sequential data of language? While embedding spaces are important, who created the linguistic structures that manifest as distributional semantics within these spaces? It is not only individual humans but humans as a collective. This question leads to the broader issue of language emergence, which is intrinsically linked with individual cognitive capabilities concerning language, including memory and attention.

Generative models, including those based on the free energy principle and active inference, have the potential to extend the discussion beyond the scope of the target paper by \citet{Parr2025beyond} to these collective questions. The Collective Predictive Coding (CPC) hypothesis \citep{Taniguchi2024collective} describes a hierarchical structure where individuals perform internal representation learning (i.e., world modeling; see \cite{Friston2021world, Taniguchi2023world}), and simultaneously, the collective performs external representation learning (i.e., language/symbol emergence; see \cite{Taniguchi2018symbol, Taniguchi2024collective}). The entire process can be characterized as (decentralized) Bayesian inference. In particular, the latter inference (external representation learning) is approximated as Bayesian inference through forms of language games, such as the Metropolis-Hastings Naming Game (e.g., \cite{taniguchi2023emergent, hoang2024emergent}).

From this perspective, an agent collective, coupled by the emergent language, can be viewed as a single entity performing active inference or predictive coding. This idea has been extended to model scientific inquiry as Collective Predictive Coding as a Model of Science \citep{Taniguchi2025cpcms}, deriving the free energy of this human collective and demonstrating its theoretical consistency with the free energy minimization of individual agents introducing a \emph{collective regularization term} representing \emph{semiotic plasticity}. Furthermore, it has been hypothesized that language emerging in this manner could exert a top-down influence on individual cognition and consciousness, such as on qualia structure \citep{Taniguchi2024constructive}.

The CPC perspective also offers an answer to the question posed by \citet{Parr2025beyond} (Section 5.2): ``Is next-word prediction the most effective way to learn language?''. We argue in the affirmative. Next-token prediction is fundamentally about modeling the probability distribution of observations, $p(\mathbf{o})$. If we adopt the CPC viewpoint that language integrates the observations of a group of partially observable agents and structurally represents $p(\mathbf{o})$, then modeling this distribution is the essence of human language learning and, through it, the learning of a collective world model \citep{Taniguchi2024generative}.

Now, let us consider the framework wherein the collective is viewed as a single super-cognitive system, termed \emph{System 3}, which extends the well-known System 1/2 framework \citep{kahneman2011thinking}, as proposed in \cite{Taniguchi2025system}. From this standpoint, perspectives on language, attention, and memory emerge that transcend the individual. For instance, through written language, society can store information externally, forming a collective memory. Individuals can also access memories stored by others through communication. Regarding attention, this super-cognitive system (the subject of group-level active inference) may be factorized into individual beings. Within their brains, the factorization in attentional mechanisms, as described by \citet{Parr2025beyond}, might then occur.

\section{Conclusion}

This commentary suggests that the discussion of memory and attention through generative models, mediated by active inference and collective predictive coding, opens avenues for inquiries extending beyond individual-level cognitive science. Such inquiries can shed more light on the mystery of language, which originates from group-level phenomena.

\bibliographystyle{plain}
\bibliography{commmentary}

\end{document}